\pdfoutput=1
\documentclass[10pt,twocolumn]{article}

\usepackage{times}
\usepackage[
    left=1.9cm,
    right=1.9cm,
    top=3.00cm,
    bottom=3.00cm
]{geometry}

\usepackage{graphicx} 
\usepackage{enumitem} 
\usepackage{xspace}
\usepackage{epsfig}
\usepackage[abbreviations]{foreign} 
\usepackage[skip=6pt, belowskip=-10pt,font=small]{caption}
\usepackage{framed}
\usepackage{url}

\usepackage{breakurl}
\usepackage[breaklinks,colorlinks=true,linkcolor=blue,urlcolor=blue,citecolor=blue]{hyperref}
\usepackage{listings}
\newcommand{\nip}[1]{\vspace{1ex}\noindent\textbf{#1}}

\newcommand {\SM} {SM\xspace}

\begin{document}

\title{Towards a Formally Verified Security Monitor for VM-based Confidential Computing}

\author{
        Wojciech Ozga\\
        \textit{IBM Research -- Zurich}\\
        \textit{woz@zurich.ibm.com}
    \and
        Guerney D. H. Hunt\\
        \textit{IBM T.J. Watson Research Center}\\
        \textit{gdhh@us.ibm.com}
    \and
        Michael V. Le\\
        \textit{IBM T.J. Watson Research Center}\\
        \textit{mvle@us.ibm.com}
    \and
        Elaine R. Palmer\\
        \textit{IBM T.J. Watson Research Center}\\
        \textit{erpalmer@us.ibm.com}
    \and
        Avraham Shinnar\\
        \textit{IBM T.J. Watson Research Center}\\
        \textit{shinnar@us.ibm.com}
}

\maketitle
\thispagestyle{plain}
\pagestyle{plain}

\begin{abstract}
Confidential computing is a key technology for isolating high-assurance applications from the large amounts of untrusted code typical in modern systems. Existing confidential computing systems cannot be certified for use in critical applications, like systems controlling critical infrastructure, hardware security modules, or aircraft, because they lack formal verification.

This paper presents an approach to formally modeling and proving a security monitor for confidential computing.  It introduces a canonical architecture for virtual machine (VM)-based confidential computing systems. It abstracts processor-specific components and identifies a minimal set of hardware primitives required by a trusted security monitor to enforce security guarantees. This paper focuses on verifying the software assuming a correct hardware implementation. 
We demonstrate our methodology and proposed approach with an example from our open-source Rust implementation of the security monitor for RISC-V.

\end{abstract}
\section{Introduction}
\label{intro}
Confidential computing, which uses terms like "enclave", "trusted execution environment", and "hardware root of trust" is designed to provide the highest protection for cloud and embedded applications. It reduces reliance on dedicated systems to protect security-critical applications.
Yet, how does one determine that a confidential computing infrastructure can be trusted? In addition to reputation, methodical design, open source development, and extensive testing, vendors rely on third party certification to assure customers that their systems and applications meet security requirements.

Applications which require the highest levels of certification are known as high assurance applications. Examples include critical infrastructure control systems~\cite{kulik2022survey}, hardware security modules (HSMs)~\cite{atsec23}, secure elements~\cite{NXPEdge}, 
and aircraft~\cite{grhillssafety}. At the highest levels of certification one of the key requirements is a formal proof of the security properties of the application and its infrastructure~\cite{smith1999validating}. We believe that confidential computing can benefit from the same rigor.

\begin{figure}[tbp!]
    \centering
    \includegraphics[width=0.48\textwidth]{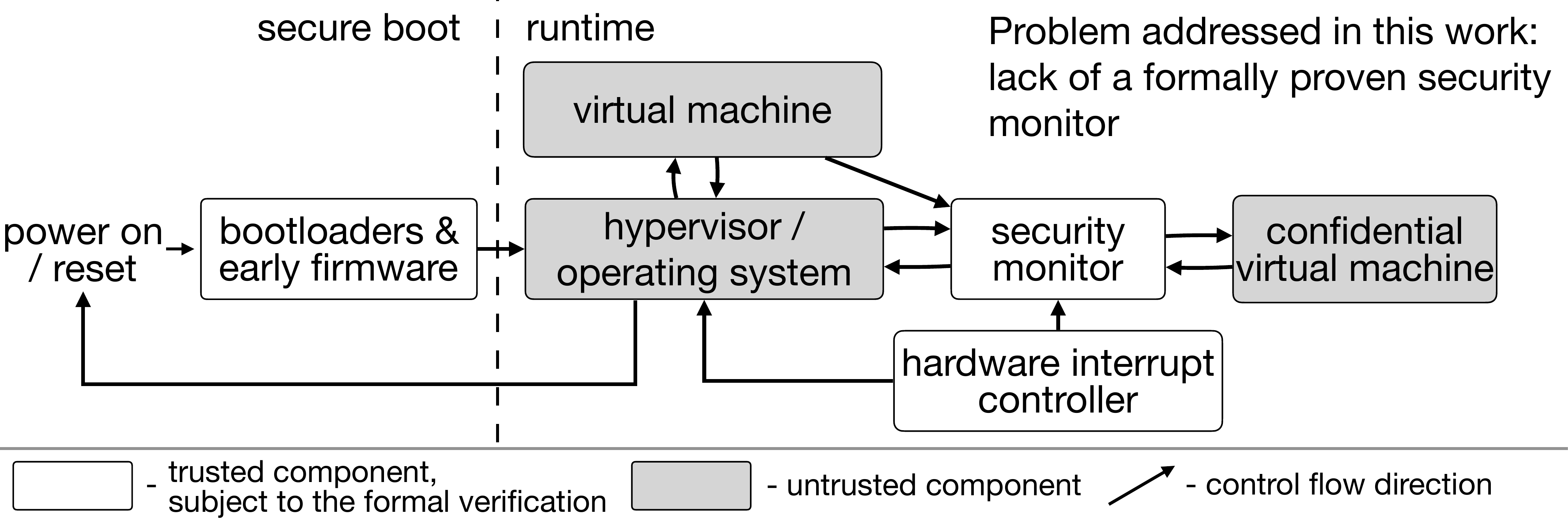}
    \caption{High-level overview of a VM-based confidential computing architecture.}
    \label{fig:overview}
    \vspace{-1ex}
\end{figure}

A key component of a confidential computing architecture is a security monitor (\SM), see \autoref{fig:overview}. 
An \SM secures confidential workloads by partitioning computing hardware in space and time. 
At a high-level, it provides a set of capabilities and guarantees to a confidential computation.
These are supported by the hardware architecture on which an \SM executes. Proving that a system is secure requires proving everything from the hardware design of the processor through the firmware. 
In addition, tools utilized to produce and verify the binary must be verified correct and uncompromised~\cite{solarwinds}.
We are not aware of any open-source, formally proven \SM for VM-based confidential computing. 

The ultimate goal of our project is to formalize, develop, and prove an open-source security monitor that can be mapped to multiple system architectures. 
This paper focuses on verifying the software with a respect to a formal model of the hardware to protect it against software attacks assuming a correct hardware implementation. The formal hardware model would provide a foundation for verifying the hardware in the future.
We start with a virtual machine-based confidential computing.   The formal model is split into architecture independent and dependent components. The architecture independent parts can be mapped to multiple confidential computing architectures with architecture dependent components that provide required invariants.
This paper introduces a part of the model of the architecture independent portion of our \SM. 

Provable systems are not new. IBM HSMs \cite{smith1999validating}, 
seL4 \cite{klein2009sel4}, and Green Hills INTEGRITY\textsuperscript{\textregistered}-178~\cite{SepKPP07,GrHills2008} are all examples of products or projects that have utilized formal methods to prove their security guarantees. 
Academic work that proves the correctness of the \SM also exists~\cite{LiARMCCA2022} but it is Arm hardware specific, so it does not provide a proof that could be used by multiple vendors. AMD~\cite{googleProjectZero2022} has utilized an independent third party review of their security monitor, but that is not a formal proof.

Our approach involves defining and formalizing a processor-independent confidential computing architecture, implementing it in Rust~\cite{matsakis2014rust}, translating the Rust implementation into Coq representation of a subset of Rust defined in separation logic~\cite{gaeher2023refinedrust, jung2017rustbelt, jung2018iris}, and proving that the implementation matches the formal model and satisfies defined security guarantees and invariants using Coq proof assistant~\cite{coq}.
There are two distinct aspects of our approach compared to existing works: 1) we define a canonical \textit{processor-independent} confidential computing architecture for which we implement the \SM and 2) we leverage the Rust memory safety properties and its ownership model to simplify the reasoning on correctness of our implementation, and hence, the proofs.

Our contributions are:
1) A methodology that enables linking and proving the design, implementation, and security guarantees of the confidential computing architecture (\S\ref{sec:methodology}).
2) A processor-independent, canonical confidential computing architecture that can be applied to a specific hardware (\S\ref{sec:overview}).
3) A part of the formal model of this architecture that lays the foundations for formal verification of the security guarantees (\S\ref{sec:model}).
4) An example based on the implementation of one of our architecture components demonstrating the proposed approach to formal verification  (\S\ref{sec:example}).

\section{Methodology}
\label{sec:methodology}
It is difficult to formally verify security guarantees of a computing system written in the natural language, such as \emph{"the security monitor guarantees confidentiality of data processed in the virtual machine's memory"}, because it is hard to mathematically express them in the context of the hardware and software architecture, operational threat model, system implementation, and its runtime behaviour. Thus, we established a methodology based on abstraction that allows us to formalize the high-level security guarantees defined in \S\ref{sec:threat_model} in a way that they can be linked to the formal model and implementation. This methodology requires the expression of all system properties and software implementation in Coq~\cite{coq}, which permits proving that the implementation conforms to the defined model, it preserves invariants and, as a result, that the security guarantees hold. 

\begin{figure}[tbp!]
\centering
\includegraphics[width=0.48\textwidth]{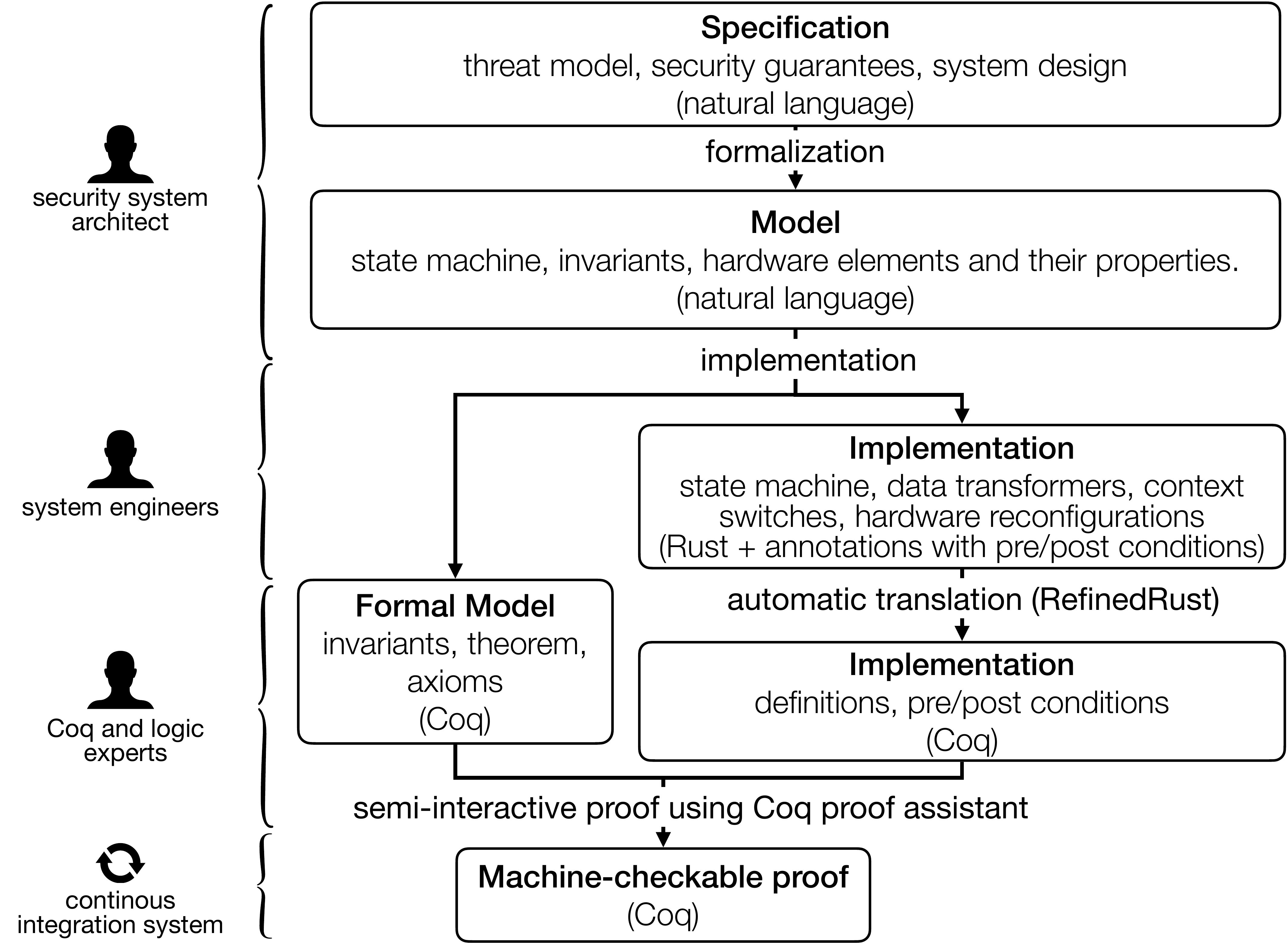}
\caption{
    Formal verification flow.
    Our approach to formal verification of the security monitor starts with the high-level specification and ends on the machine-checkable proof that the implementation implements the model and preserves invariants required to maintain the security guarantees.
}
\label{fig:workflow}
\end{figure}
\autoref{fig:workflow} shows an overview of our approach. We define the system architecture on several levels of abstraction (high-level specification, architecture, formal model, implementation) that omit lower level details while becoming more and more specialized until the lowest level at which we have formal definitions of the model and implementation expressed with mathematical formulas.

The formal verification requires building the mathematical model of the system design and showing that the implementation translates to the formal model (completeness) and preserves invariants. To reason about the soundness of the implementation, we must prove its functional correctness which requires proving memory and execution safety. To simplify the reasoning about the memory safety, we rely on the type system of the Rust programming language~\cite{matsakis2014rust} that is based on ownership, borrowing, and lifetimes.  

An important question from practitioners is how to prove the correctness of the system design and derived formal model itself, which was used to prove the system. Indeed skilled individuals are still needed to specify the invariants of the system and verify that the threat model is covered. Since we cannot formally verify the high-level specification of a formal model expressed in natural language, we require that auditors with expert knowledge in the field of computer architecture and systems security analyze its soundness. This is the current state-of-the-art approach in the field of confidential computing where the public specification of the architecture, sometimes including the source code of the security model, is analyzed by experts~\cite{cheng2023intel, aktas2023tdx, googleProjectZero2022}. 
In our methodology, the auditors must also verify that the definitions of the specification and the formal model have been correctly formalized in Coq. To gain additional assurance, we can also verify that our formalization satisfies various expected properties. After this point, the rest of the proof can be automated because the specification and implementation are defined using formal mathematical expression, such as axioms, theorems, pre/post conditions, and logical sentences. As is common in other works, we rely on correctness of formal verification tooling~\cite{klein2009sel4, gaeher2023refinedrust}. Our methodology requires placing trust in the formalization of Rust, i.e., $\lambda_{Rust}$~\cite{jung2017rustbelt}, the Coq's kernel proof checker~\cite{coq}, and translation of the Rust intermediary representation to machine code.
	
Our approach to the formal verification differs from other approaches because we implement the \SM in Rust, a programming language that provides memory safety~\cite{ballo2023rustsafety} 
and a type system that can enforce object ownership. This allows us (with minor exceptions requiring extra proof efforts) not to tackle the problem of memory accesses using pointer arithmetic that are root of safety and security problems in unsafe languages like C and assembly. Trying to formally verify these unsafe languages leads to state explosion, requiring the use of, for example, reduced programming.  We can also take advantage of Rust's object ownership regime to encode and enforce additional properties, such as exclusive access to memory regions, as discussed in Section~\ref{sec:example}.

Our methodology is modular. We omit the verification of the functional correctness of certain components, i.e., hardware, instead assuming they work properly and offer properties expressed in our formal model as axioms. The formal verification of the hardware design and implementation (e.g.,~\cite{cheang2022verifying}) must be performed separately to complete the proof of the end system. 
\section{Architecture Overview} 
\label{sec:overview}
\begin{figure}[tbp!]
\centering
\includegraphics[width=0.48\textwidth]{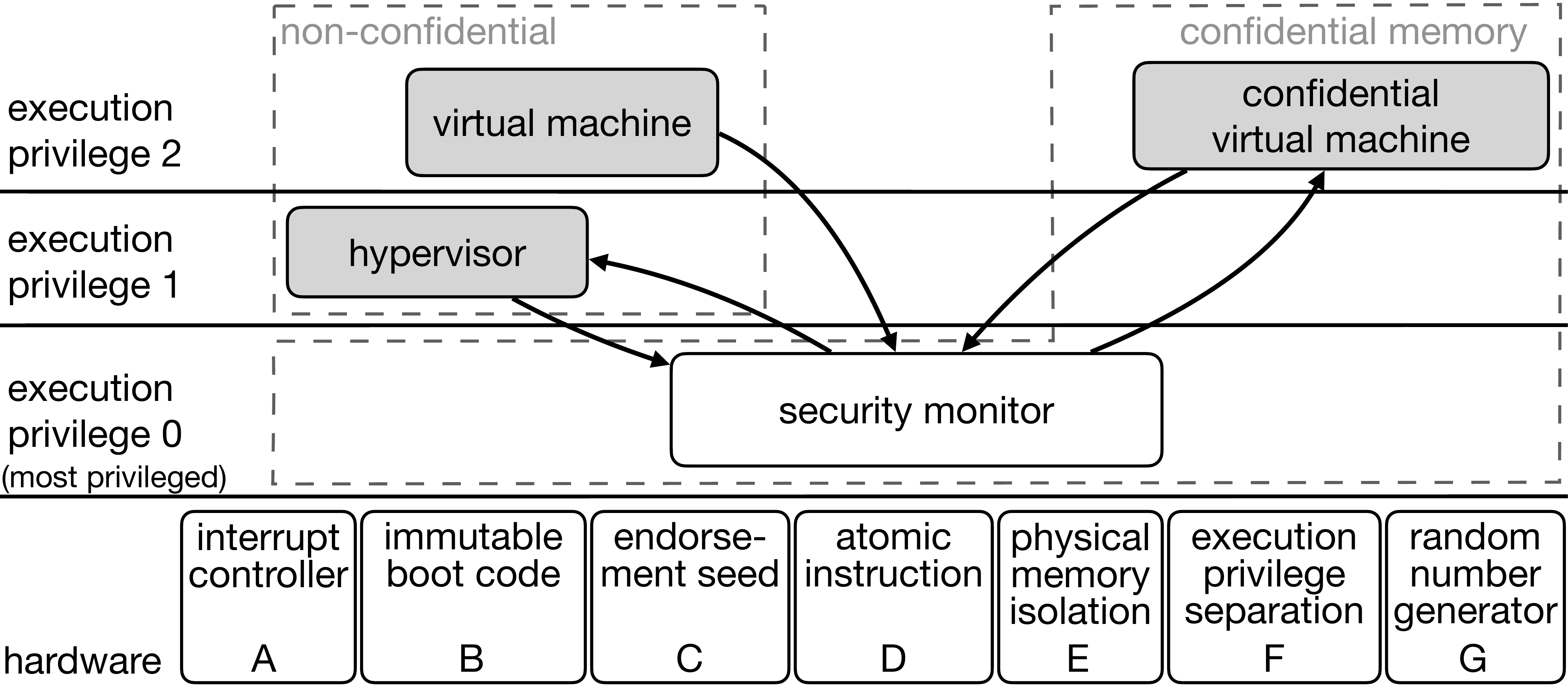}
\caption{Canonical confidential computing architecture in the context of required hardware and software components.}
\label{fig:archandflow}
\end{figure}

\autoref{fig:archandflow} shows a high-level overview of an architecture enabling confidential computing requiring minimal hardware support.
It consists of four groups of components: 
(1) a set of seven hardware components that enable isolating computations and memory,
(2) the \SM that enforces the security guarantees of the confidential workload,
(3) an untrusted hypervisor that manages the life-cycle of virtual machines, and 
(4) legacy virtual machines (VMs) that execute non-confidential workload and confidential virtual machines (CVMs) that run confidential workload.

\nip{Hardware functionalities:}
The interrupt controller (A) enables signaling and passing control flow between execution privileges. It presents interrupts directly to the hypervisor or the \SM. There is at least one interrupt that always is presented to the \SM. Interrupts that normally go to the hypervisor can be reconfigured to go to the \SM. 
However, the reconfiguration of an interrupt targeted to the highest execution privilege is restricted to the \SM.

The immutable boot code (B) enables integrity-and authenticity-enforced boot of the security monitor. This code is where execution of the system starts with interrupts disabled at power on or after a reset. This code is the core root of trust (CRTM) in the architecture. Prior to exiting to the next level code the hardware must be  configured so that the CRTM is immutable. This code implements the secure and trusted boot~\cite{arbaugh1997secureboot}. 

The endorsement seed (C) is required for attestation. It is a secret random number that enables the derivation of an attestation key used to sign measurements and information that certifies the manufacturer of the system. 

At least one atomic instruction (D) is required on multi-core processors to construct synchronization primitives, such as spin lock, mutex, or semaphore.

Physical memory isolation (E) allows isolating memory regions by setting and enforcing memory access control. This component supports marking memory as confidential or non-confidential. CVMs can request that they are granted non-confidential memory, which is initialized to zero. A non-confidential thread\footnote{Thread --- instructions executed by a processor at certain execution privilege.} cannot access confidential memory and a confidential thread can only access non-confidential memory that has been returned as a result of a sharing request.

Execution privilege separation (F) enables partitioning software to create, assign, and enforce roles and their access control. It also supports marking threads as either confidential or non-confidential. Marking memory or a thread as confidential is restricted to the highest privilege level.

Random number generator (G) enables security monitor to obtain random numbers required for cryptographical operations.

\nip{Security Monitor (SM):}
The \SM runs at the highest execution privilege.
Its main runtime functions are call routing, state transformation, and hardware reconfiguration. 
The \SM receives request from VMs, the hypervisor, and CVMs.

The \SM partitions memory into confidential and non-confidential.
The four main areas of confidential memory are: 1) \SM region, 2) control data region, 3) confidential virtual machines region, and 4) uninitialized confidential memory region. The \SM region contains the code, data, stack, and heap. 
The control data region is used by the \SM to preserve the processor state associated with the process that was interrupted for it to receive control. 
The confidential virtual machine region is where the CVM's code and data are located. 
The \SM utilizes the hardware to guarantee that 
the \SM and control data regions can only be referenced by the \SM. It also has to guarantee that each CVM can only reference the memory region containing its own code and data.
The \SM does not offer protections for shared non-confidential memory.

\nip{Untrusted Software:} 
The hypervisor and all non-confidential VMs execute in the same security domain. The hypervisor provides services to VMs and CVMs. The \SM controls the transition from CVMs to the hypervisor and from the hypervisor to CVMs. When a CVM makes a hypervisor call, the \SM decides whether to preform the call or route the call to the hypervisor. When a call is routed, the \SM shields the state of the CVM from the hypervisor except for what is required to preform the call. When the call returns, the \SM returns to the CVM only the responses from the hypervisor, shielding the hypervisor state from the CVM. 

\nip{CVMs:}
All CVMs start out as a VM and call the \SM to transition to a CVM. After moving the VM to secure memory, the \SM calculates and saves the measurement of the virtual machine.
The \SM provides an attestation call to enable the owner of the new CVM instance to cryptographically verify the hardware support and the instance of the CVM\footnote{Our high level model supports both local \cite{hunt2021pef} and remote attestation.}.

Prior to transferring control to a CVM, the \SM re-configures the system to receive all interrupts. This is to maintain control on context switches between different security domains during which extra steps are required to enforce the security guarantees.

\subsection{Threat Model}
\label{sec:threat_model}
The canonical architecture assumes a software-level adversary who controls all untrusted software including the hypervisor, OS, user-space software, VMs and some confidential VMs (CVMs), excluding the victim CVM.
The attacker's goal is to (1) read confidential data of the victim CVM, (2) force the victim CVM to process the data of the adversary's choice, (3) change the execution flow of the victim CVM, or (4) impersonate the victim CVM to its owner/user.
An attacker can start, stop, interrupt the victim CVM at arbitrary point in time. She can provide arbitrary input data via virtual I/O devices, registers, and shared memory buffers. An attacker can also interact with peripheral devices not assigned to a confidential VM. Protection against covert channels and side-channels between security domains is addressed by the architecture. However, an in-depth discussion is elided for space reasons.

The canonical architecture excludes denial of service, physical attacks on processor, memory, or buses. It can be extended, as has been done by several architectures, with cryptographic protection of data leaving the processor to protect against this vector of attacks. For example, RISC-V includes a concept of per-HART encrypted memory.
The processor detects faults, recovers from faults or stops execution (i.e., we do not cover fault injection except for faults corrected by hardware, like error correction codes (ECC)).

\subsection{Security Guarantees}
\label{sec:sec_guarantees}

The design of our architecture provides mechanisms that can be used to guarantee confidentiality\footnote{Similar to PEF~\cite{hunt2021pef}, we require that the disk of the VM be encrypted, and that the key for decryption is not inserted until after attestation. Therefore the code executed before attestation only has integrity protection and not confidentiality.} and integrity of the data and the code of a CVM, including the runtime state (content of the processor registers and cache) and data offloaded to processor-external storage (main memory).
This guarantee can only be achieved if the CVM is itself correct and secure, which the CVM owner must ensure. 
The architecture also enables the user of the CVM to verify using attestation that the code executing in the CVM is the expected code and that it is executing under a trusted security infrastructure. 

The \SM and boot code are trusted.
By design, the CVM has no communication/data flow to other code executing in the system in other security domains. 
During CVM execution, the CVM must explicitly enable such flows, e.g., by declaring a portion of memory to be shared with the hypervisor and using it as a communication buffer, for example, for VirtIO.

The \SM enforces the following four security policies: 1) data isolation (data belonging to a security domain remains private and not accessible by other security domains) including data isolation of the non-confidential domain from confidential domains, 2) control of information flow (information flow originated from authorised sources), 3) sanitization of resources (prevent revealing information after context switches, e.g., via registers or micro-architectural state), and 4) fault isolation prevention (failures do not cascade across security domains). 

\section{Model}
\label{sec:model}

\subsection{Execution Environment}
\label{sec:exec_env}
Our canonical confidential computing architecture relies on functions provided by \emph{security-critical hardware components} (\autoref{fig:archandflow} A-F). To define an architecture-independent security monitor (\SM), we extracted the required hardware properties and defined them as axioms:

\nip{A.HW.1:} The hardware offers at least 3 execution privileges.
The software executing in the highest execution privilege is the only component of the system permitted to reconfigure security-critical hardware components. 

\nip{A.HW.2:}
A hardware mechanism protects the initial boot code from untrusted modifications. A processor reset results in the execution of the initial boot code. The first instruction executes at the highest execution privilege. 

\nip{A.HW.3:} The memory isolation component can deny accesses originated from the processor and peripheral devices to a defined set of physical memory addresses (corresponding to confidential memory).

\nip{A.HW.4:} The processor offers a way to clear all the micro-architectural state leaving no traces of the previous execution.

\nip{A.HW.5:}  The interrupt controller delivers interrupts to the defined trap handlers and performs a context switch to the assigned execution privilege. Interrupts are disabled when presented to the trap handler. Trap handlers are responsible to enable interrupts prior exit. The interrupt controller enables the interrupts to be configured as enabled or disabled. Only code running at the highest execution privilege can target or reconfigure an interrupt to execute at this privilege. 

\nip{A.HW.6:} Either the processor exposes an endorsement seed after reset and software can enable a lock that prevents access to the seed until the next processor reset, or such an endorsement seed is securely provided by an external component.

\nip{A.HW.7:} Processor provides an atomic instruction for reading and writing memory.

\nip{A.HW.8:} Either the processor exposes a random number generator or such generator is securely provided by an external component.

Axioms $A.HW.1$ and $A.HW.2$ define an environment where isolation of the \SM and security-critical hardware from all untrusted software can be accomplished.  
Otherwise there are no guarantees that the hardware properties are preserved in the presence of untrusted software and peripheral devices. 
Axiom $A.HW.3$ defines a way to isolate regions of physical memory from all software and hardware components of the system. Otherwise, other software could easily overwrite the \SM's code or data.
Axiom $A.HW.4$ guarantees that there is a way to hide the execution traces between concurrent execution of different security domains.
Axiom $A.HW.5$ defines requirement on the interrupt controller which when not under the \SM's control could be used by the untrusted software to take control of the execution at arbitrary point of time and read the confidential data via micro-architectural state.
Axiom $A.HW.6$ defines the endorsement seed as a unique, secret random number with hardware support to make it inaccessible until the next reboot. Otherwise untrusted software could use the secret to impersonate the \SM during attestation.
Axiom $A.HW.7$ defines that the processor offers atomic instructions required for implementing synchronization primitives. Otherwise, we could not guarantee \SM's data integrity on multi-core systems.
Finally, $A.HW.8$ defines a random number generator (RNG) that is required for cryptographic operations. Without the RNG, a CVM could not create a secure communication channel with the verifier because it could neither create its own key nor generate nonces.

The OpenPOWER/POWER9 is an example of existing hardware that provides the required hardware properties~\cite{hunt2021pef}. Active work is underway in the RISC-V hardware architecture community to fulfill all these properties.

\subsection{Initialization}
\label{sec:initialization}
The formal model requires that after the processor reset and before execution of untrusted code, the \SM controls the entire computing environment and re-configures it in a way to guarantee its own isolation. We state these guarantees as:

\nip{S.Init.1:} No other software or hardware component but the \SM can modify the \SM's code or data or modify the security-critical hardware configuration.

\nip{S.Init.2:} Only the \SM can access the endorsement seed and the derived attestation key.

To satisfy these requirement, we rely on the secure boot~\cite{arbaugh1997secureboot} that bootstraps the computer from immutable code ($A.HW.2$).
This piece of code leads to the execution of the initialization procedure that sets up the system in a way that the following invariants are preserved until the next processor reset, i.e., during the entire life cycle of the \SM:

\nip{I.Init.1:} The \SM executes entirely and exclusively at the highest execution privilege. Any other security domain, i.e., hypervisor or CVMs, cannot execute at this privilege level.

\nip{I.Init.2:} The memory isolation component controls access to the confidential memory.
The confidential memory is accessible only by code executing from the confidential memory.

\nip{I.Init.3:} The code and data of the \SM resides entirely in the confidential memory.

\nip{I.Init.4:} The interrupt controller is configured in a way that all (at least one) interrupts that must be handled at the highest execution privilege are configured to be handled by the \SM.

\nip{I.Init.5:} Access to the endorsement seed is read and write protected and the derived attestation key is stored in the control data region.

The invariants above guarantee $S.Init.1$ and $S.Init.2$. After the trusted code boots up the platform ($A.HW.2$) it passes control to the \SM's initialization procedure in the highest privilege mode. This procedure 1) isolates its code and data ($I.Init.2$ and $I.Init.3$) from untrusted code that will execute in lower execution privileges ($A.HW.1$ and $A.HW.3$), 2), makes sure that only the \SM handles interrupts at the highest execution privilege ($A.HW.5$ and $I.Init.4$), 3) removes access to the endorsement seed and the attestation key ($A.HW.6$ and $I.Init.5$), 4) clears all micro-architectural state ($A.HW.4$), and 5) enables interrupts and passes control to untrusted software at a lower execution privilege  ($A.HW.1$ and $I.Init.1$).

Because we require secure and measured boot of all firmware prior to OS execution the fact that a secure monitor initialization procedure executed before any untrusted code it verifiable later using an attestation protocol. If the signed measurements are not provided by a separate device~\cite{will2015tpm}, it can be produced using the attestation key derived from the endorsement seed~\cite{tcg2020dice}. Since this seed and key are accessible only by the \SM ($A.HW.6$ and $I.Init.5$), no one can impersonate.

\subsection{Runtime Finite State Machine}
\label{sec:fsm}
\autoref{fig:fsm} shows the finite state machine (FSM) representing the execution flow of the security monitor during runtime, i.e., after the platform initialization.

The security monitor's job is to respond to calls by determining the state transformation it is requested to execute and deciding if such transformation does not violate the promised security guarantees. If the request does not violate guarantees  the security monitor will apply that transformation, and eventually return execution.

The FSM has two parts, the non-confidential (NC) and confidential (C) part. The NC part contains nodes that process requests from the hypervisor/VM. The C part contains nodes that process requests from the CVM. The security monitor exits to the hypervisor only from the NC-exit node and exits to the CVM only from the C-exit node. Transitions $NC \rightarrow C$ or $C \rightarrow NC$ indicate incoming transition between security domains and, therefore, must apply state sanitization and protection mechanisms, e.g., re-configuration of the memory isolation component or restore of the execution state at correct entry point. The FSM maintains the following invariants:

\begin{figure}[tbp!]
    \centering
    \includegraphics[width=0.48\textwidth]{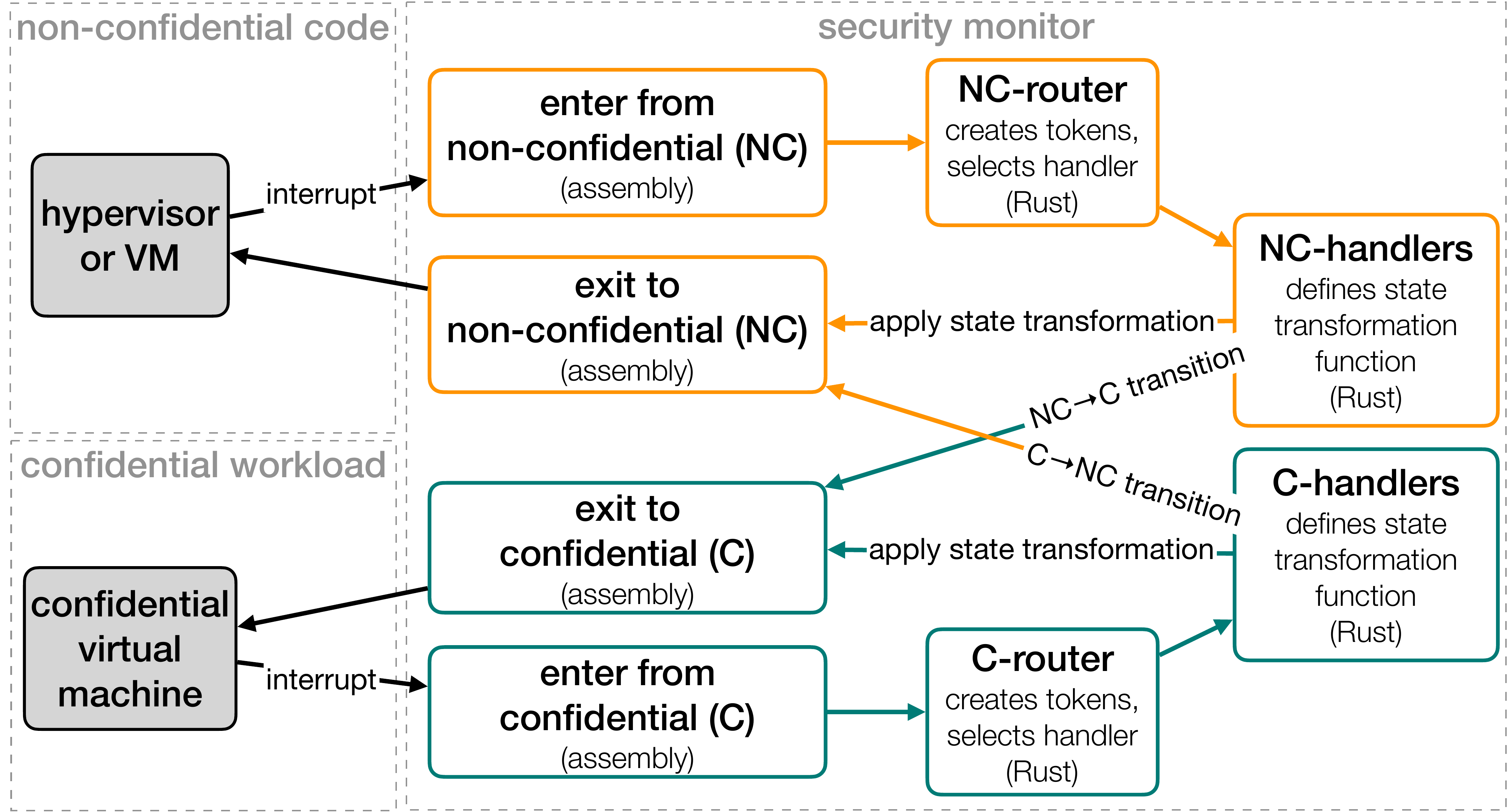}
    \caption{State machine representing the security monitor's control flow execution.}
    \label{fig:fsm}
\end{figure}

\nip{I.FSM.1:} The security monitor runs with interrupts disabled.

\nip{I.FSM.2:} $NC \rightarrow C$ transition results in: 1) re-configuring the memory isolation component so that the CVM has access to its code and data located in confidential memory and has no access to non-confidential memory (unless explicitly requested), 2) saves the state and interrupt delivery configuration of the hypervisor or VM, 3) reconfigures all interrupts being delivered to the security monitor. 

\nip{I.FSM.3:} $C \rightarrow NC$ transition results in: 1) re-configuring memory isolation component to deny access to confidential memory, 2) restore of the state and  interrupt delivery configuration to what was saved during the $NC \rightarrow C$ transistion. 

\nip{I.FSM.4:} Enter nodes result in saving the security domain's processor state in the confidential memory's control data region.

\nip{I.FSM.5:} Exit nodes clear micro-architectural state and restore the security domain's processor state from the confidential memory's control data region.

\nip{I.FSM.6:} Except for the narrow interface defined by the state transformation (see below), the architectural state of one security domain is never visible to another security domain. 

Initially the security monitor provides the full-isolation of the confidential workload, preventing any form of communication with other security domains. However, this approach is not practical because the CVM could not share the results of its computation or communicate with other security domains.
We provide an \SM call that results in a creation of a shared non-confidential memory page between the CVM and the hypervisor. This shared page could be used to contain a communications buffer used by CVM's VirtIO to communicate with the hypervisor.

The \SM performs state transformation which restricts the information passed from the CVM to the hypervisor to only the information required to preform the request. It also restricts the information returned to the CVM to only the results of the request. The exact information enabled during state transformation is architecture dependent. Additional details on the finite state machine are elided for space reasons.

More advanced architectures might further relax some of these invariants to improve performance with the help of more sophisticated hardware. For example, a security monitor might allow direct interrupt delegation to the CVM. Additional efforts would be required then for the formal verification of the interrupt delegation configuration and interrupt controller correctness.
\section {Implementation}
\label{sec:implementation}

\nip{Hardware.}
We implemented a security monitor (\SM) that supports the canonical VM-based confidential computing architecture in Rust and targets RISC-V. The \SM implementation is open source\footnote{\url{https://github.com/IBM/ACE-RISCV}}. To meet the hardware requirements defined by our model, we exploited different RISC-V technologies which provide properties defined as axioms in \S\ref{sec:exec_env}.

We build on top of the RISC-V 64-bit processor implementing the atomic (A) and hypervisor (HS) extensions because it provides the required number of privilege modes ($A.HW.1$), atomic instructions that allow us to implement synchronization primitives based on the spinlock ($A.HW.7$), required instructions for clearing micro-architectural state ($A.HW.4$).

The RISC-V physical memory protection (PMP) technology \cite{waterman2021riscvpriv}, IOPMP \cite{ku2023iopmp}, and 2-level page table address translation provide required properties to isolate confidential memory ($A.HW.3$). PMP/IOPMP have limitation on number of regions they can isolate. Direct hardware support for a single hardware providing equivalent properties is ongoing work within the RISC-V community~\cite{sahita2023smmtt}.

We rely on the core-local interrupt controller (CLIC) and platform-level interrupt controller (PLIC) as local and external interrupt controllers ($A.HW.5$). The future version of virtualization-aware and confidential computing-aware PLIC could be used to increase performance by directly delivering interrupts to the target CVM.

At the moment there is no ratified RISC-V specification defining requirements for secure boot and access to the endorsement seed ($A.HW.2$ and $A.HW.6$). The RISC-V Security Model TG is working on the specification which we plan to adapt.

\nip{Software.}
We use OpenSBI~\cite{opensbi} as the boot and SBI firmware~\cite{riscv_sbi}. The OpenSBI invokes the \SM's initialization procedure and  passes control to the Linux kernel-based hypervisor. During the initialization, the \SM assigns pages for the processor save areas, heap, splits memory into confidential and non-confidential parts using PMP, and re-configures the interrupt delegation, so all HS- and VS-ecalls trap in the \SM's context switch handler. 

We implemented the \SM in Rust according to the finite state machine defined in \S\ref{sec:fsm}. 
The context switches are implemented in assembly and perform store and load of the processor's state to and from the confidential memory. 

The current version of the \SM supports initialization and transformations that enable: conversion of the VM to CVM, resume and termination of the CVM, sharing a memory page between the hypervisor and CVM, exposing a limited number of hypercalls, and enabling VirtIO by exposing memory-mapped I/O (MMIO) load/store requests. 
\section {Example: Memory Tracker}
\label{sec:example}
\begin{figure}[tbp!]
    \centering
    \includegraphics[width=0.48\textwidth]{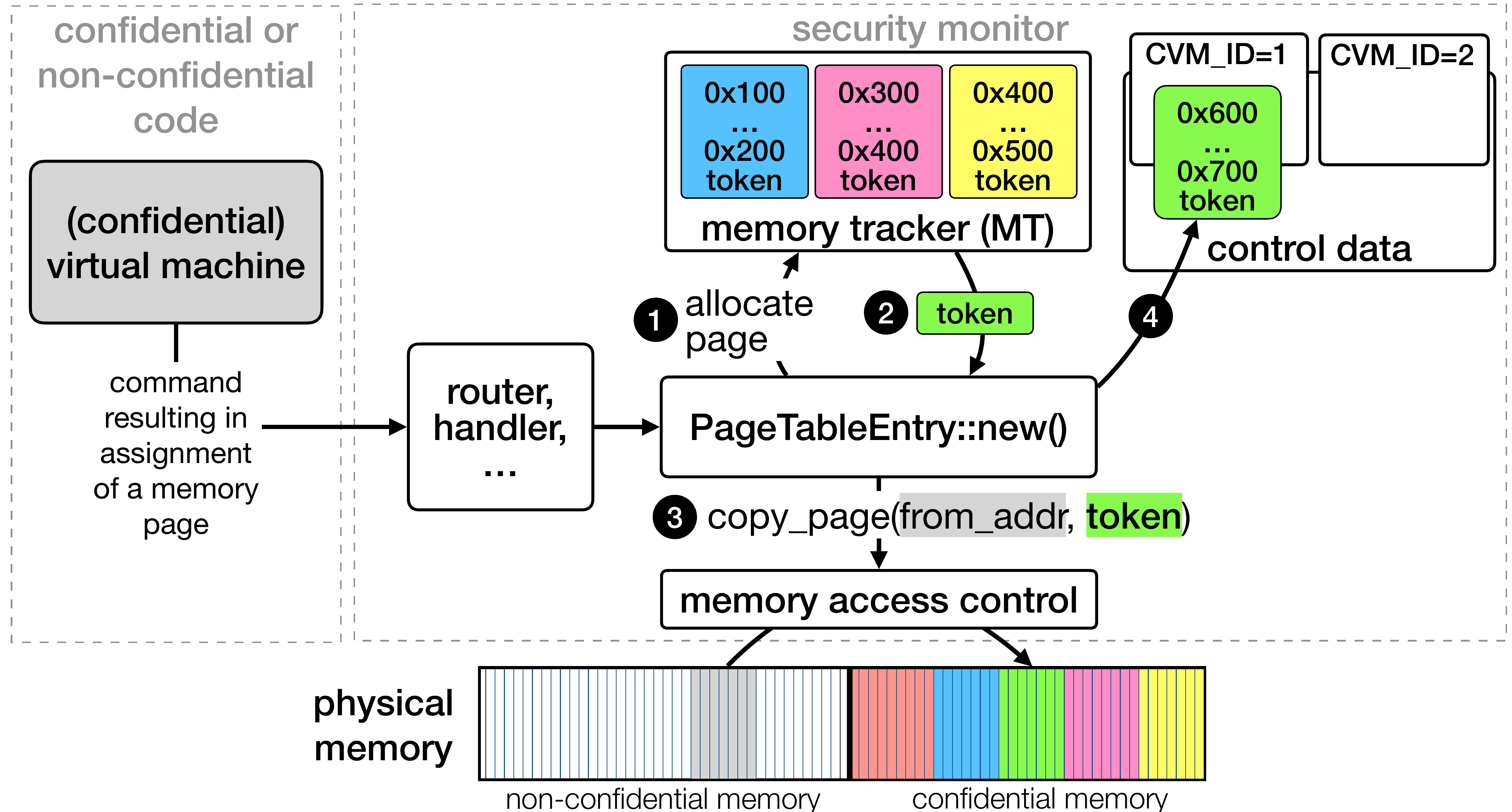}
    \caption{Memory tracker manages confidential memory pages to enforce that a single page cannot be assigned to two different CVMs.}
    \label{fig:mem_tracker}
\end{figure}

To provide an example of our methodology, we discuss the implementation and the proof direction of one of the \SM's critical components. \autoref{fig:mem_tracker} presents the memory tracker, an internal module of the \SM that allocates chunks of confidential memory at page size granularity. The main goal of the memory tracker is to help ensure confidentiality invariants. The security guarantee provided by the memory tracker are as follows: 

\nip{S.MT.1:} The memory tracker enforces that two disjoint security domains, e.g., two different confidential VMs, cannot own the same physical memory region.

\subsection{Tokens}
To simplify the proof of this claim, we leverage Rust's ownership model and its memory safety. We introduce the concept of a token called \emph{Page}, which represents the aligned physical memory page. The component of the \SM that owns (in the Rust meaning) the instance of the Page logically owns the corresponding physical memory region defined by the token. Carefully creating a single token per page simplifies reasoning about the runtime memory allocation because safe Rust enforces that a single token can only be transferred but not duplicated or owned by two different entities simultaneously.

This example shows a conceptual use of tokens and consists of three components: (1) the MemoryTracker component (\autoref{lst:memory_tracker}) that owns multiple (2) Page tokens (\autoref{lst:page}) and gives them away to (3) PageTable (\autoref{lst:page_table}). The PageTable component configures the  memory management unit to provide CVMs with virtual memory. Tokens are to prevent cases where entries in page tables belonging to different CVMs would point to the same memory region, i.e., breaking the $S.MT.1$ guarantee.

\begin{lstlisting}[caption=The Page token component (Rust),label={lst:page}, language=c,breaklines=true,breakatwhitespace=true,escapechar=@]
pub trait PageState {}
pub enum UnAllocated {}
pub enum Allocated {}
impl PageState for UnAllocated {}
impl PageState for Allocated {}

pub struct Page<S: PageState> { @\label{lst:page:page_def}@
  address: usize,
  size: PageSize,
  _marker: PhantomData<S>,
}

impl Page<UnAllocated> { @\label{lst:page:unallocated}@
  pub(super) fn init(address: usize, size: PageSize) -> Self {@\label{lst:page:constr}@
    Self { address, size, _marker: PhantomData }
  }
  pub fn zeroize(self) -> Page<Allocated> { @\label{lst:page:transition1}@
    self.clear();
    Page { address: self.address, size: self.size, _marker: PhantomData } @\label{lst:page:t1}@
  }
}

impl Page<Allocated> { @\label{lst:page:allocated}@
  pub fn deallocate(self) -> Page<UnAllocated> { @\label{lst:page:transition2}@
    self.clear();
    Page { address: self.address, size: self.size, _marker: PhantomData } @\label{lst:page:t2}@
  }
}
\end{lstlisting}

\autoref{lst:page} shows the conceptual definition of the Page token (line \ref{lst:page:page_def}), which describes a memory region starting at a certain address and having a specific size. The Page token can be in two states distinguishable by the Rust type system: UnAllocated and Allocated (lines \ref{lst:page:unallocated} and \ref{lst:page:allocated}). The Rust's type system enforces that only an UnAllocated Page can be created using the constructor (line \ref{lst:page:constr}) and transformed to an Allocated Page after clearing the corresponding memory region (line \ref{lst:page:transition1}). Deallocating the page, a process required before allocating the page to a different security domain, requires transitioning the token to the UnAllocated state because this is the type accepted by the memory tracker (\autoref{lst:memory_tracker} line \ref{lst:mt:vec}). Deallocation is only possible by calling a function that clears the content of the corresponding memory region (line \ref{lst:page:transition2}), a process that gives additional assurance that confidential data is correctly isolated in time. 

\begin{lstlisting}[caption=The MemoryTracker component (Rust),label={lst:memory_tracker}, language=c,breaklines=true,breakatwhitespace=true,escapechar=@]
pub struct MemoryTracker { @\label{lst:mt:def}@
  pages: Vec<Page<UnAllocated>>, @\label{lst:mt:vec}@
}

impl MemoryTracker {
  pub fn allocate(&mut self) -> Option<Page<UnAllocated>> {@\label{lst:mt:alloc}@
    self.pages.pop()
  }

  pub fn deallocate(&mut self, mut pages: Vec<Page<Allocated>>) { @\label{lst:mt:dealloc}@
    let p: Vec<_> = pages.drain(..).map(|f| f.deallocate()).collect();
    self.pages.append(&mut p);
  }
}
\end{lstlisting}

\autoref{lst:memory_tracker} shows the definition of the conceptual memory tracker component (line \ref{lst:mt:def}). It initially owns all unallocated page tokens and stores them inside a vector (line \ref{lst:mt:vec}). 
Note that the Rust's type system prevents the memory tracker from storing the Allocated Page tokens that already belong to some component, e.g., a page table of some CVM.
The only ways for the safe Rust code to access a token is by requesting it from the memory tracker using the allocate() function (\autoref{lst:memory_tracker} line \ref{lst:mt:alloc}) or by calling the token constructor (\autoref{lst:page} line \ref{lst:page:constr}). We ensure that the constructor is called only once by the boot code. Everywhere else tokens can only be transferred via the allocate() function.

\begin{lstlisting}[caption=The PageTable component (Rust),label={lst:page_table}, language=c,breaklines=true,breakatwhitespace=true,escapechar=@]
pub struct PageTable { @\label{lst:pt:def}@
  page: Page<Allocated>, @\label{lst:pt:allocated}@
}

impl PageTable {
  pub fn empty(unallocated_page: Page<UnAllocated>) -> Result<Self, Error> { @\label{lst:pt:unallocated}@
    let page = unallocated_page.zeroize(); @\label{lst:pt:conversion}@
    Ok(Self { page }) @\label{lst:pt:created}@
  }

  pub fn set_entry(&mut self, index: usize, entry: usize) {@\label{lst:pt:set_entries}@
    if index < MAX_NUMBER_OF_ENTRIES {
      let offset = index * size_of::<usize>();
      self.page.write(offset, entry);
    }
  }
}

impl Page<Allocated> { @\label{lst:pt:alloc}@
  pub fn write(&mut self, offset: usize, value: usize) { @\label{lst:pt:entries}@
    if offset+size_of::<usize>() < self.size.in_bytes() {
      let paddr = self.address + offset;
      unsafe { @\label{lst:pt:us_start}@
        (paddr as *mut usize).write_volatile(value);
      }
    }
  }
}
\end{lstlisting}

Listing \ref{lst:page_table} shows definition of a PageTable, which is an \SM's component for configuring the memory management unit's  page tables (line \ref{lst:pt:def}). The PageTable instance can only be created from the UnAllocated Page token (line \ref{lst:pt:unallocated}) that is converted to an Allocated Page when creating the empty (no virtual memory mappings yet) page table (lines \ref{lst:pt:conversion} and \ref{lst:pt:created}). Access to page table entries (line \ref{lst:pt:set_entries}), i.e., content of the physical memory, requires use of unsafe Rust (line \ref{lst:pt:us_start}). However, it is encapsulated in a safe interface (line \ref{lst:pt:entries}) of the Allocated Page token (line \ref{lst:pt:alloc}) that prevents writing to the memory outside the range defined by the token. The correctness of the implementation using unsafe Rust requires extra proving efforts~\cite{gaeher2023refinedrust, matsushita2022rusthornbelt}.

\subsection{Invariants}
We define the following invariants that, once hold, prove the security guarantee $S.MT.1$ as shown in the next section.

\nip{I.MT.1:} The fixed set of Page tokens is created only by the initialization function. Allocating or deallocating a token does not increase/decrease the size of the set.
The initialization function can be invoked only during the secure boot process.

\nip{I.MT.2:} The initialization function creates one Page token per aligned physical page in a way that every pair of Page tokens defines disjoint physical memory regions.

\nip{I.MT.3:} A read or write to a memory location requires owning the associated Page token.

\vspace{1ex}
The invariant $I.MT.1$ guarantees that after the initialization there is a fixed number of tokens and no additional tokens can be created later. This prevents generation of arbitrary tokens during runtime. The invariant $I.MT.2$ states that the functionally correct algorithm generates tokens which do not define overlapping memory regions. Otherwise, there could be two tokens defining the same memory address, opening up a possibility for two different security domains unwillingly sharing the same memory address. The invariant $I.MT.3$ is to guarantee the immutability of tokens in the face of code not obeying safe Rust rules. 

\subsection{Implementation and Proof directions}
Our proof of $S.MT.1$ starts with the assumption that
there exist two page tables defining the same physical memory address in different security domains and shows a contradiction.
This means that each of the PageTables owns a Page token which defines the same physical memory address. This requires that (1) both PageTables use the same Page token, or (2) two different Page tokens describe overlapping memory addresses. (1) is false because of the Rust ownership, its memory safety guarantees, and the invariant $I.MT.3$. (2) is false because of invariants $I.MT.1$ and $I.MT.2$. Thus, the proposition leads to the contradiction which completes the proof.

The formal verification of the implementation must show that invariants $I.MT.1$ - $I.MT.3$ hold during runtime because only then does the above proof holds. We can show that the invariant $I.MT.1$ holds by proving that the memory tracker's init() constructor is the only code of the \SM which creates Pages and is referenced only from the initialization function called by the boot firmware. One could show this with a simple \emph{grep} tool or using Rust's scopes that limit visibility of the function within the \SM, like \emph{pub(super)} in \autoref{lst:page} line \ref{lst:page:constr}.

Proving that invariant $I.MT.2$ holds requires proving the functional correctness of the memory tracker's constructor that creates all pages. Such a proof requires translating the code into a formalized Rust representation, e.g., a $\lambda_{Rust}$ \cite{jung2017rustbelt}, and proving it with the Coq proof assistant~\cite{coq}.

Proving the invariant $I.MT.3$ requires more complete Rust representation that covers unsafe Rust. This is provided by RefinedRust~\cite{gaeher2023refinedrust} which allows annotating code with pre/post conditions and compiling the code to formalized representation of Rust based on $\lambda_{Rust}$ and, eventually, proving it with the Coq proof assistant.
\section{Related Work}
There is a rich history as well as ongoing work in applying formal methods to improve the safety~\cite{grant2020review} and security of systems~\cite{kulik2022survey}. 
Works that advocate for designing confidential computing architectures that are open, modular, and hardware agnostic also exist~\cite{opentee_2020, lee2020keystone, sanctum, ferraiuolo2017komodo}.
As our work is at the intersection of architectural design and formal verification, we focus the below discussion on works that generalize confidential computing architecture and/or verify aspects of confidential computing along with their toolings.

\nip{Komodo}~\cite{ferraiuolo2017komodo} 
demonstrated the effectiveness of separating the fundamental security mechanisms of confidential computing hardware from the management of it. A formal specification (in Dafny~\cite{leino2010dafny}) verified the correctness and security properties of the \SM.

\nip{Arm CCA} is a VM-based confidential computing architecture for Arm. Its \SM is verified in~\cite{LiARMCCA2022} to meet a design specification expressed in Coq along with its associated security properties.

\nip{Keystone}~\cite{lee2020keystone} proposes a framework for building modular customizable trusted execution environments using unmodified hardware, even without explicit RISC-V support for them. A small, privileged \SM manages security boundaries enforced by hardware primitives. Keystone does not have a formal specification. 

\nip{The seL4 microkernel} was the "first formal proof of functional correctness of a complete, general-purpose operating system kernel." \cite{klein2009sel4} Modeling a general-purpose operating system, even a smaller microkernel, is a significantly larger and more difficult effort than modeling an \SM with supporting hardware.

\nip{Information flow analysis on HDL}~\cite{ifc_hdl_2017} proposes the use of static information flow analysis to verify processor designs expressed in a hardware description language. They demonstrated their approach by implementing and verifying a prototype of a simple multi-core Arm TrustZone architecture. 

\nip{Islaris}~\cite{sammler2022islaris} is a system to verify machine code using real-world specifications (Armv8-A and RISC-V) written in the Sail ISA definition language and modeled in Coq. This work does not need to trust the compiler, but does have to deal with pointer arithmetic. 


\vspace{1ex}
Our project focus and methodology shares much with the above work especially with Komodo~\cite{ferraiuolo2017komodo}, Keystone~\cite{lee2020keystone} and Arm CCA's \SM verification~\cite{LiARMCCA2022} with key differences:
1) we focus on VM-based rather than process-based SGX-like confidential computing design, 2) our design is canonical and architecturally independent, and 3) we leverage properties of the Rust language to simplify the correctness reasoning of our implementation, and hence, the proofs.

\section {Conclusion and Future Work}
We introduced a processor-independent, canonical VM-based confidential computing architecture and a methodology to prove it. We formalized the architecture, defining the required hardware properties, initialization procedure, and an \SM. We provide a proof outline of a security property of an internal component of the \SM implementation for RISC-V.

Future work will focus on formal modelling of all components of our architecture, proving them using Coq, and refining our canonical architecture for RISC-V implementation. For that we will compile our Rust implementation to Coq and prove it using the Coq proof assistant. At the same time, we plan to add extensions to the canonical architecture that will enable use of hypercalls required to run the full Linux-based operating system, enable cryptographic protection of confidential memory, and leverage the dynamic allocation of confidential memory with the RISC-V Smmtt. Completing this project will show that the Rust language has sufficient support to formally verifying a security monitor.

Presented project is open source and we invite collaborators to work together to push the boundaries of confidential computing technology. The open-source implementation of the security monitor is publicly available: \url{https://github.com/IBM/ACE-RISCV}.

\bibliographystyle{ieee}
\bibliography{bibliography}

\begin{thebibliography}{10}

\bibitem{aktas2023tdx}
{\sc Aktas, E., Cohen, C., Eads, J., Forshaw, J., and Wilhelm, F.}
\newblock {Intel Trust Domain Extensions (TDX) Security Review}.
\newblock {\em Google technical report\/} (2023).

\bibitem{arbaugh1997secureboot}
{\sc Arbaugh, W.~A., Farber, D.~J., and Smith, J.~M.}
\newblock A secure and reliable bootstrap architecture.
\newblock In {\em Proceedings. 1997 IEEE Symposium on Security and Privacy
  (Cat. No. 97CB36097)\/} (1997), IEEE, pp.~65--71.

\bibitem{will2015tpm}
{\sc Arthur, W., and Challener, D.}
\newblock {\em A {Practical} {Guide} to {TPM} 2.0: {Using} the {Trusted}
  {Platform} {Module} in the {New} {Age} of {Security}}.
\newblock Apress, 2015.

\bibitem{atsec23}
{\sc atsec information~security corporation}.
\newblock {IBM 4769-001 Enterprise PKCS\#11 HSM Cryptographic Coprocessor
  Security Module, FIPS 140-2 Non-Proprietary Security Policy}, Jul 2023.

\bibitem{ballo2023rustsafety}
{\sc Ballo, T.}
\newblock {Blue Team Rust: What is "Memory Safety", Really?}
\newblock \url{https://tiemoko.com/blog/blue-team-rust/}, accessed on August
  2023.

\bibitem{cheang2022verifying}
{\sc Cheang, K., Rasmussen, C., Lee, D., Kohlbrenner, D.~W., Asanovi{\'c}, K.,
  and Seshia, S.~A.}
\newblock Verifying risc-v physical memory protection.
\newblock {\em arXiv preprint arXiv:2211.02179\/} (2022).

\bibitem{cheng2023intel}
{\sc Cheng, P.-C., Ozga, W., Valdez, E., Ahmed, S., Gu, Z., Jamjoom, H.,
  Franke, H., and Bottomley, J.}
\newblock Intel tdx demystified: A top-down approach.
\newblock {\em arXiv preprint arXiv:2303.15540\/} (2023).

\bibitem{googleProjectZero2022}
{\sc Cohen, C., Forshaw, J., Horn, J., and Brand, M.}
\newblock Amd secure processor for confidential computing security review.
\newblock Technical report, Google Project Zero and Google Cloud Security, May
  2022.
\newblock Accessed early 2023.

\bibitem{sanctum}
{\sc Costan, V., Lebedev, I., and Devadas, S.}
\newblock Sanctum: Minimal hardware extensions for strong software isolation.
\newblock In {\em 25th USENIX Security Symposium (USENIX Security 16)\/}
  (Austin, TX, Aug. 2016), pp.~857--874.

\bibitem{ferraiuolo2017komodo}
{\sc Ferraiuolo, A., Baumann, A., Hawblitzel, C., and Parno, B.}
\newblock {Komodo: Using verification to disentangle secure-enclave hardware
  from software}.
\newblock In {\em 26th ACM Symposium on Operating Systems Principles\/}
  (October 2017), ACM, pp.~287--305.
\newblock The Komodo specification, prototype implementation, and proofs are
  available at https://github.com/Microsoft/Komodo.

\bibitem{ifc_hdl_2017}
{\sc Ferraiuolo, A., Xu, R., Zhang, D., Myers, A.~C., and Suh, G.~E.}
\newblock Verification of a practical hardware security architecture through
  static information flow analysis.
\newblock In {\em International Conference on Architectural Support for
  Programming Languages and Operating Systems (ASPLOS)\/} (2017).

\bibitem{gaeher2023refinedrust}
{\sc Gaeher, L., Sammler, M., Jung, R., Krebbers, R., and Dreyer, D.}
\newblock {RefinedRust: Towards high-assurance verification of unsafe Rust
  programs}.
\newblock In {\em Rust Verification Workshop\/} (2023).

\bibitem{grant2020review}
{\sc Grant, E.~S., and Nanda, S.~P.}
\newblock A review of applications of formal specification in safety-critical
  system development.
\newblock In {\em Proceedings of the 2020 the 4th International Conference on
  Compute and Data Analysis\/} (2020), pp.~208--215.

\bibitem{grhillssafety}
{\sc {Green Hills Software}}.
\newblock {INTEGRITY\textregistered-178 tuMP RTOS Safety-Critical \&
  Security-Critical RTOS}.
\newblock
  https://www.ghs.com/products/safety\_critical/integrity\_178\_certifications.html.
\newblock accessed 14 Aug 2023.

\bibitem{tcg2020dice}
{\sc Group, T.~C.}
\newblock {DICE} attestation architecture.
\newblock In {\em TCG Resources\/} (2020).

\bibitem{hunt2021pef}
{\sc Hunt, G. D.~H., Pai, R., Le, M.~V., Jamjoom, H., Bhattiprolu, S., Boivie,
  R., Dufour, L., Frey, B., Kapur, M., Goldman, K.~A., Grimm, R., Janakirman,
  J., Ludden, J.~M., Mackerras, P., May, C., Palmer, E.~R., Rao, B.~B., Roy,
  L., Starke, W.~A., Stuecheli, J., Valdez, E., and Voigt, W.}
\newblock {Confidential Computing for OpenPOWER}.
\newblock In {\em Proceedings of the Sixteenth European Conference on Computer
  Systems\/} (2021), EuroSys '21.

\bibitem{jung2017rustbelt}
{\sc Jung, R., Jourdan, J.-H., Krebbers, R., and Dreyer, D.}
\newblock Rustbelt: Securing the foundations of the rust programming language.
\newblock {\em Proc. ACM Program. Lang.\/} (2017).

\bibitem{jung2018iris}
{\sc Jung, R., Krebbers, R., Jourdan, J.-H., Bizjak, A., Birkedal, L., and
  Dreyer, D.}
\newblock Iris from the ground up: A modular foundation for higher-order
  concurrent separation logic.
\newblock {\em Journal of Functional Programming 28\/} (2018), e20.

\bibitem{klein2009sel4}
{\sc Klein, G., Elphinstone, K., Heiser, G., Andronick, J., Cock, D., Derrin,
  P., Elkaduwe, D., Engelhardt, K., Kolanski, R., Norrish, M., et~al.}
\newblock sel4: Formal verification of an os kernel.
\newblock In {\em Proceedings of the ACM SIGOPS 22nd symposium on Operating
  systems principles\/} (2009), pp.~207--220.

\bibitem{opentee_2020}
{\sc Kohlbrenner, D., Shinde, S., Lee, D., Asanovic, K., and Song, D.}
\newblock Building open trusted execution environments.
\newblock {\em IEEE Security \& Privacy 18}, 05 (sep 2020), 47--56.

\bibitem{ku2023iopmp}
{\sc Ku, P., and Tang, C.}
\newblock {RISC-V IOPMP Specification Document: version 1.0.0-draft1}.
\newblock
  \url{https://github.com/riscv-admin/iopmp/blob/main/specification/riscv_iopmp_specification.pdf},
  accessed on August 2023.

\bibitem{kulik2022survey}
{\sc Kulik, T., Dongol, B., Larsen, P.~G., Macedo, H.~D., Schneider, S.,
  Tran-J{\o}rgensen, P.~W., and Woodcock, J.}
\newblock A survey of practical formal methods for security.
\newblock {\em Formal Aspects of Computing 34}, 1 (2022), 1--39.

\bibitem{lee2020keystone}
{\sc Lee, D., Kohlbrenner, D., Shinde, S., Asanovi{\'c}, K., and Song, D.}
\newblock Keystone: An open framework for architecting trusted execution
  environments.
\newblock In {\em Proceedings of the Fifteenth European Conference on Computer
  Systems\/} (2020), pp.~1--16.
\newblock Repository: https://github.com/orgs/keystone-enclave/repositories.

\bibitem{leino2010dafny}
{\sc Leino, K. R.~M.}
\newblock Dafny: An automatic program verifier for functional correctness.
\newblock In {\em International conference on logic for programming artificial
  intelligence and reasoning\/} (2010), Springer, pp.~348--370.

\bibitem{LiARMCCA2022}
{\sc Li, X., Li, X., Dall, C., Gu, R., Nieh, J., Sait, Y., and Stockwell, G.}
\newblock Design and verification of the arm confidential compute architecture.
\newblock In {\em 16th USENIX Symposium on Operating Systems Design and
  Implementation (OSDI 22)\/} (Carlsbad, CA, July 2022), USENIX Association,
  pp.~465--484.

\bibitem{matsakis2014rust}
{\sc Matsakis, N.~D., and Klock, II, F.~S.}
\newblock The rust language.
\newblock In {\em Proceedings of HILT\/} (2014).

\bibitem{matsushita2022rusthornbelt}
{\sc Matsushita, Y., Denis, X., Jourdan, J.-H., and Dreyer, D.}
\newblock Rusthornbelt: a semantic foundation for functional verification of
  rust programs with unsafe code.
\newblock In {\em Proceedings of the 43rd ACM SIGPLAN International Conference
  on Programming Language Design and Implementation\/} (2022), pp.~841--856.

\bibitem{SepKPP07}
{\sc {National Security Agency Information Assurance Directorate}}.
\newblock {U.S. Government Protection Profile for Separation Kernels in
  Environments Requiring High Robustness, v1.03}, 2007.

\bibitem{NXPEdge}
{\sc {NXP Semiconductors}}.
\newblock {Certified EdgeLock\textsuperscript{\textregistered} Assurance}.
\newblock
  \url{https://www.nxp.com/products/nxp-product-information/nxp-product-programs/edgelock-assurance:EDGELOCK-ASSURANCE},
  accessed on August 2023.

\bibitem{opensbi}
{\sc {RISC-V International, Western Digital Corporation or its affiliates}}.
\newblock {RISC-V Open Source Supervisor Binary Interface (OpenSBI)}.
\newblock \url{https://github.com/riscv-software-src/opensbi}, accessed on
  August 2023.

\bibitem{riscv_sbi}
{\sc {RISC-V Platform Runtime Services Task Group}}.
\newblock {RISC-V Supervisor Binary Interface Specification, version 2.0-rc2}.
\newblock
  \url{https://github.com/riscv-non-isa/riscv-sbi-doc/blob/master/riscv-sbi.pdf},
  accessed on August 2023.

\bibitem{sahita2023smmtt}
{\sc Sahita, R., Dellow, A., Liberty, D., Gupta, D., Hunt, G., Asanovic, K.,
  Hill, M., Wood, N., Koyuncu, O., Elliott, P., Ortiz, Samuel abd~Shanbhogue,
  V., and Ozga, W.}
\newblock {RISC-V SmmTT: Supervisor Domain Access Protection}.
\newblock \url{https://github.com/riscv/riscv-smmtt/}, accessed on August 2023.

\bibitem{GrHills2008}
{\sc {SAIC, Inc.}}
\newblock {Common Criteria Evaluation and Validation Scheme Validation Report,
  Green Hills Software INTEGRITY-178B Separation Kernel}, 2008.

\bibitem{sammler2022islaris}
{\sc Sammler, M., Hammond, A., Lepigre, R., Campbell, B., Pichon-Pharabod, J.,
  Dreyer, D., Garg, D., and Sewell, P.}
\newblock Islaris: verification of machine code against authoritative isa
  semantics.
\newblock In {\em Proceedings of the 43rd ACM SIGPLAN International Conference
  on Programming Language Design and Implementation\/} (2022), pp.~825--840.
\newblock Repository: https://github.com/rems-project/islaris.

\bibitem{SamFinger22}
{\sc {Samsung Electronics Co., Ltd.}}
\newblock {Samsung Introduces Smart All-in-One Fingerprint Security IC for
  Biometric Payment Cards}.
\newblock
  \url{https://www.businesswire.com/news/home/20220124005881/en/Samsung-Introduces-Smart-All-in-One-Fingerprint-Security-IC-for-Biometric-Payment-Cards},
  accessed on August 2023.

\bibitem{selfridge2019grift}
{\sc Selfridge, B.}
\newblock Grift: A richly-typed, deeply-embedded risc-v semantics written in
  haskell.
\newblock In {\em Proc. SpISA 2019: Workshop Instr. Set Architect.
  Specification\/} (2019).
\newblock Repository: https://github.com/GaloisInc/grift.

\bibitem{smith1999validating}
{\sc Smith, S., Perez, R., Weingart, S., and Austel, V.}
\newblock Validating a high-performance, programmable secure coprocessor.
\newblock In {\em Proceedings, 22nd National Information Systems Security
  Conference\/} (1999).

\bibitem{solarwinds}
{\sc Temple-Raston, D.}
\newblock A 'worst nightmare' cyberattack: The untold story of the solarwinds
  hack.
\newblock
  \url{https://www.npr.org/2021/04/16/985439655/a-worst-nightmare-cyberattack-the-untold-story-of-the-solarwinds-hack},
  accessed on August 2023.

\bibitem{coq}
{\sc {The Coq Team}}.
\newblock {The Coq proof assistant}.
\newblock \url{https://coq.inria.fr/}, accessed on August 2023.

\bibitem{waterman2021riscvpriv}
{\sc Waterman, A., Asanovi, K., and Hauser, J.}
\newblock {The RISC-V Instruction Set Manual, Volume II: Privileged
  Architecture, Document Version 20211203}.
\newblock
  \url{https://github.com/riscv/riscv-isa-manual/releases/download/Priv-v1.12/riscv-privileged-20211203.pdf},
  accessed on August 2023.

\end{thebibliography}

\end{document}